\documentclass[preprint,1p]{elsarticle}
\usepackage{amsmath,amsfonts,amssymb,graphics,graphicx,epsfig,color,times}
\usepackage{graphics}
\usepackage[utf8x]{inputenc}
\usepackage{verbatim}
\usepackage{color}
\usepackage{bbm} 
\usepackage{subfigure}
\usepackage{hyperref}
\usepackage{mathrsfs}
\journal{Annals of Physics}
\begin{document}
\begin{frontmatter}
\title{Cooperative single-photon subradiant states in a three-dimensional atomic array}
\author{H. H. Jen}
\ead{sappyjen@gmail.com}
\address{Institute of Physics, Academia Sinica, Taipei 11529, Taiwan, R. O. C.}
\begin{abstract}
We propose a complete superradiant and subradiant states that can be manipulated and prepared in a three-dimensional atomic array.\ These subradiant states can be realized by absorbing a single photon and imprinting the spatially-dependent phases on the atomic system.\ We find that the collective decay rates and associated cooperative Lamb shifts are highly dependent on the phases we manage to imprint, and the subradiant state of long lifetime can be found for various lattice spacings and atom numbers.\ We also investigate both optically thin and thick atomic arrays, which can serve for systematic studies of super- and sub-radiance.\ Our proposal offers an alternative scheme for quantum memory of light in a three-dimensional array of two-level atoms, which is applicable and potentially advantageous in quantum information processing. 
\end{abstract}
\begin{keyword}
Subradiance; Superradiance; Atomic array; Single photon fluorescence; Quantum memory
\end{keyword}
\end{frontmatter}

\section{Introduction}
Superradiance \cite{Dicke1954} and associated collective phenomena \cite{Friedberg1973, Gross1982, Mandel1995} have raised continuous interests in the past sixty years.\ These collective effects are due to the induced dipole-dipole interaction \cite{Stephen1964, Lehmberg1970} that comes from the common field mediating the atomic system.\ This resonant dipole-dipole interaction is in essence long-ranged, therefore the decay behavior heavily depends on the interatomic distance and the geometry of the atomic ensemble.\ The spontaneous decay can exhibit superradiance that is more commonly observed in experiments, and also cooperative Lamb shift (CLS) \cite{Friedberg1973, Scully2009} and subradiance, that are less observable because of the demanding precision and signal-to-noise ratio in measurements.

Recent studies involve a fast decay of second-order correlation of two photons from the cascade atomic ensemble \cite{Chaneliere2006, Jen2012, Srivathsan2013}, and a redshift of CLS in the embedded Fe atoms in the planar cavity \cite{Rohlsberger2010}, the atomic vapor layer \cite{Keaveney2012}, an ionic atomic system \cite{Meir2014}, and a cold-atom ensemble \cite{Friedberg2010, Pellegrino2014, Jen2015}.\ Subradiant decay can also be measured in diversified atomic systems of ring/disk plasmonic nanocavities \cite{Sonnefraud2010}, ultracold molecules \cite{McGuyer2015}, and a large cloud of cold atoms \cite{Guerin2016}.\ Interestingly, recent proposals of singly-excited states that can describe single-photon superradiance \cite{Scully2006, Eberly2006, Mazets2007, Svidzinsky2008} and subradiance \cite{Scully2015, Vetter2016, Jen2016} provide a new direction in investigating these collective effects in a limited but complete Hilbert space.\ The advantage of the setting of single-photon interacting with the atoms restricts this space to $N$ singly-excited states and one ground state, hugely simplifying the dynamical light-matter interacting systems of for example $2^N$ states of $N$ two-level atoms.

In this paper we propose a complete Hilbert space of cooperative single-photon states that are responsible for the superradiance and subradiance.\ In section 2, we introduce these states that can be prepared in a three-dimensional (3D) atomic array with one atom per site.\ In section 3, we introduce the theoretical background of our analysis.\ We then investigate the time evolutions of the subradiant states in section 4, which can be observable in fluorescence experiments, and we discuss and conclude in section 5.

\section{Cooperative single-photon states using De Moivre's formula}
When a single photon interacts with two-level atoms of $|g\rangle$ and $|e\rangle$ for the ground and excited states respectively, the so-called timed Dicke state is formed on absorbing this photon \cite{Scully2006},
\begin{eqnarray}
|\phi_N\rangle=\frac{1}{\sqrt{N}}\sum_{\mu=1}^N e^{i\mathbf{k}\cdot \mathbf{r}_\mu}|e\rangle_\mu|g\rangle^{\otimes(N-1)},
\end{eqnarray}
where $\mathbf{k}$ is the wavevector of single photon.\ The above denotes a superposition of one and only one excited state with the rest of ($N-1$) ground state atoms.\ This timed Dicke state is symmetrical under exchange of any two atoms, thus shows superradiant decay after the absorption.\ The dipole-dipole interaction is responsible for the emergence of enhanced decay behavior.\ However since this symmetrical state is not the eigenstate for this long-ranged dipole-dipole interaction in general, it couples with other nonsymmetrical (NS) and orthogonal states during the process of spontaneous emission.\ These NS states on the other hand are responsible for both super- and subradiant decays, and can be observed in the emission long after the superradiant decay \cite{Mazets2007, Guerin2016}.\ The possible candidates for these NS states can be found in Refs. \cite{Mazets2007, Svidzinsky2008, Scully2015, Vetter2016, Jen2016}.\ Following the proposal of NS states in one-dimensional atomic array \cite{Jen2016}, we further extend it to a 3D case which is more advantageous in the efficiency of absorbing the photon and has richer physics in super- and sub-radiance due to its dimensionality.

\begin{figure}[t]
\centering
\includegraphics[width=10cm,height=7cm]{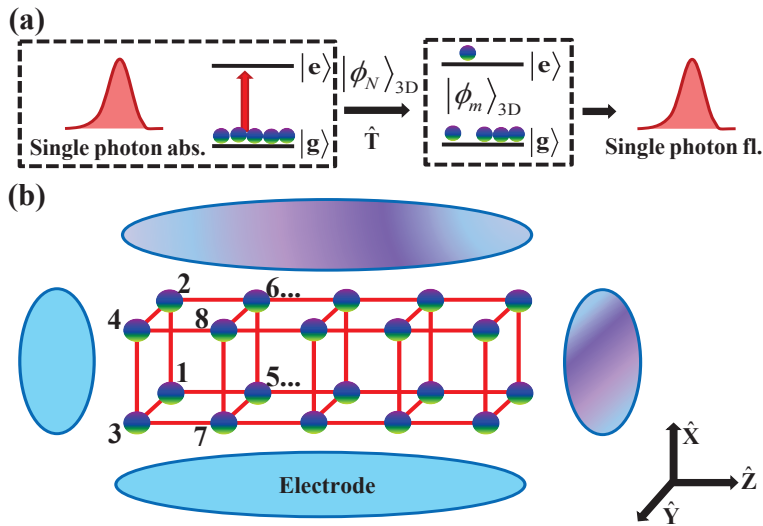}
\caption{(Color online) Schematic preparation of DM states in a 3D atomic array.\ (a) A symmetric state $|\phi_N\rangle_{\rm 3D}$ is created with a single photon absorption (abs.), which evolves to other super- or sub-radiant DM states $|\phi_m\rangle_{\rm 3D}$ via a unitary transformation $\hat{T}$ to imprint the required phases.\ Later single-photon fluorescence (fl.) measurements confirm the DM state preparation. (b) The required spatially-varying phases can be imprinted via electrodes along three axes (we omit the electrodes along $\hat{y}$ for clear demonstration), and the labeling of atomic indices is shown on the side of the array as an example of the ordering.}\label{fig1}
\end{figure}

In \cite{Jen2016}, we utilize the De Moivre's formula to construct the complete Hilbert space of singly-excited states and the extension to a 3D atomic array is straightforward,
\begin{eqnarray}
|\phi_m\rangle_{\rm 3D}=\sum_{\mu=1}^N \frac{e^{i\mathbf{k}\cdot\mathbf{r}_\mu}}{\sqrt{N}}e^{i\frac{2m\pi}{N}(\mu-1)}|e\rangle_\mu|g\rangle^{\otimes(N-1)},
\end{eqnarray}
where we denote them as DM (De Moivre) states and their normalizations are ensured.\ The orthonormality can be shown as 
\begin{eqnarray}
_{\rm 3D}\langle\phi_m|\phi_n\rangle_{\rm 3D}=\frac{1}{N}\sum_{\mu=1}^Ne^{i\frac{2\pi}{N}(\mu-1)(m-n)} =\delta_{m,n}.
\end{eqnarray}
The De Moivre's formula is intentionally used for finding the nth root of unity, where there exists a complex number $z$ such that $z^n$ $=$ $1$.\ For the construction of $N$ singly-excited states, we then take the roots of $z^N$ $=$ $1$ as state coefficients which are $e^{i2\pi m/N}$ for $m$ $\in$ [$1$ , $N$] in the above.\ The extra phases (other than the light propagating phase $e^{i\mathbf{k}\cdot\mathbf{r}_\mu}$) of the above singly-excited states can be imprinted by a Zeeman or Stark field gradient which is commonly used in controlled reversible inhomogeneous broadening (CRIB) \cite{Kraus2006} in quantum memory of light \cite{Hetet2008, Hedges2010, Sparkes2010, Hosseini2011}.\ The labeling of the imprinted phases associated with the atomic indices $\mu$'s however is not arbitrary.\ For the discrete and linearly increasing phases we manipulate to imprint on the atomic array, these $\mu$'s need to be labeled also linearly in specific orders along the axes.\ The order of the axes matters only on the relative gradient strength applied, and without loss of generality we choose the labeling order first along $\hat{x}$, then $\hat{y}$ and $\hat{z}$ as shown in figure \ref{fig1}.

For schematic demonstration, in figure \ref{fig1} we show the proposed setting of investigating super- and sub-radiance in a 3D atomic array.\ The preparation of the DM states $|\phi_m\rangle_{\rm 3D}$ can be done by first preparing the symmetrical state $|\phi_N\rangle_{\rm 3D}$ on absorbing a single photon.\ Then we evolve $|\phi_N\rangle_{\rm 3D}$ to $|\phi_m\rangle_{\rm 3D}$ via a unitary transformation to imprint the required linearly increasing phases.\ The phases can be accumulated either by a Zeeman field gradient \cite{Jen2016}, an ac Stark field utilizing beam shaping \cite{Sparkes2010}, or electrodes used in praseodymium dopants in yttrium orthosilicate (Pr$^{3+}$:Y$_2$SiO$_5$) \cite{Hedges2010}.\ Therefore the unitary transformation is $\hat{T}$ $\equiv$ $e^{iV_{ph}\tau}$, where the interaction energy $V_{ph}$ could be $-\mu\cdot B$ from the magnetic field gradient $B'$ $=$ $B/z$ or $-d\cdot E$ from Stark field gradient $E'$ $=$ $E/z$, along the $\hat{z}$ axis for example.\ The interaction time is $\tau$ which determines the amount of the phases for specific DM state.

We note that the amplitudes of the coefficients are equal in our proposed DM states which are thus endowed with a cooperative nature.\ In the next section we introduce the theoretical background for the analysis of the property and the time evolutions of these DM states we manage to prepare.

\section{Theoretical background of induced dipole-dipole interaction}
The dynamical time evolution of a single photon interacting with $N$ two-level atoms can be described by the Hamiltonian ($V_{\rm I}$) in an interaction picture ($\hbar$ $=$ $1$) \cite{Lehmberg1970},
\begin{eqnarray}
V_{\rm I}=-\sum_{\mu=1}^{ N}\sum_{\mathbf{k},\lambda} g_{\mathbf{k}}(\epsilon_{\mathbf{k},\lambda}\cdot\hat{d})\hat{S}_\mu[e^{-i(w_{\mathbf{k}}t-\mathbf{k}\cdot\mathbf{r}_\mu)}\hat{a}_{\mathbf{k},\lambda}+\rm h. c.] \label{V},\nonumber\\
\end{eqnarray}
and the dipole operator is defined as 
\begin{eqnarray}
\hat{S}_\mu\equiv\hat{\sigma}_\mu e^{-i\omega_{eg}t}+\hat{\sigma}_\mu^\dag e^{i\omega_{eg}t},
\end{eqnarray}
where $\hat{\sigma}_\mu$ $\equiv$ $|g\rangle_\mu\langle e|$ and the transition frequency is $\omega_{eg}$ $=$ $\omega_e$ $-$ $\omega_g$.\ The coupling constant of light-matter interaction is $g_{\mathbf{k}}$, light polarization is $\epsilon_{\mathbf{k},\lambda}$, and the unit direction of the dipole operator is $\hat{d}$.\ The above is a typical treatment for a quantized bosonic field ($\hat{a}_{\mathbf{k},\lambda}$) interacting with the atoms using the dipole approximation \cite{QO:Scully}, while the counter rotating-wave terms are kept in order to correctly account for the frequency shift of the dipole-dipole interaction which we show below.

Solving Heisenberg equations of motion from the above Hamiltonian with secular approximations \cite{Lehmberg1970}, we derive the time evolution for arbitrary atomic operators $\hat{Q}$ in a Lindblad form that
\begin{eqnarray}
\frac{\dot{\hat{Q}}}{\Gamma}=i\sum_{\mu\neq\nu}^{ N}\sum_{\nu=1}^{ N} G_{\mu\nu}\left[\hat{\sigma}_{\mu}^\dag\hat{\sigma}_\nu,\hat{Q}\right]
+\sum_{\mu=1}^{ N}\sum_{\nu=1}^{ N} F_{\mu\nu}\left[\hat{\sigma}_{\mu}^\dag\hat{Q}\hat{\sigma}_\nu-\frac{1}{2}\left(\hat{\sigma}_{\mu}^\dag\hat{\sigma}_\nu\hat{Q}
+\hat{Q}\hat{\sigma}_{\mu}^\dag\hat{\sigma}_\nu\right)\right],
\end{eqnarray}
where $F_{\alpha,\beta}$ and $G_{\alpha,\beta}$ are defined as \cite{Lehmberg1970}
\begin{eqnarray}
F_{\mu\nu}(\xi)&\equiv&
\frac{3}{2}\bigg\{\left[1-(\hat{d}\cdot\hat{r}_{\mu\nu})^2\right]\frac{\sin\xi}{\xi}+\left[1-3(\hat{d}\cdot\hat{r}_{\mu\nu})^2\right]\left(\frac{\cos\xi}{\xi^2}-\frac{\sin\xi}{\xi^3}\right)\bigg\},\\
G_{\mu\nu}(\xi)&\equiv&\frac{3}{4}\bigg\{-\Big[1-(\hat{d}\cdot\hat{r}_{\mu\nu})^2\Big]\frac{\cos\xi}{\xi}
+\Big[1-3(\hat{d}\cdot\hat{r}_{\mu\nu})^2\Big]
\left(\frac{\sin\xi}{\xi^2}+\frac{\cos\xi}{\xi^3}\right)\bigg\}.
\end{eqnarray}
The dimensionless length scale is $\xi$ $=$ $|\mathbf{k}| r_{\mu\nu}$, and the relative distance is $r_{\mu\nu}$ $=$ $|\mathbf{r}_\mu-\mathbf{r}_\nu|$ with the transition wave number $|\mathbf{k}|$.\ Single particle (intrinsic) decay constant is $\Gamma$ as a frequency scale.\ The above is the origin of resonant dipole-dipole interaction induced by the common light-matter interaction and rescattering events in the interacting medium.\ $F_{\mu\nu}$ and $G_{\mu\nu}$ are essentially long-ranged, and respectively they are cooperative decay rates and cooperative Lamb shifts.

To investigate the time evolution of DM states in a 3D atomic array, we use the Schr\"{o}dinger equations projected from the above Lindblad form.\ First we define $|\psi_\mu\rangle$ $\equiv$ $|e\rangle_\mu|g\rangle^{(N-1)}$ as the bare state bases, and their time evolutions can be derived by using $\hat{Q}$ $=$ $|\psi_\mu\rangle\langle g|^{\otimes N}$ projecting on $|g\rangle^{\otimes N}$.\ Then the state of the atomic system in Schr\"{o}dinger picture can be expressed as $|\Psi(t)\rangle$ $=$ $\sum_{\mu=1}^N c_\mu(t)|\psi_\mu\rangle$, where the probability amplitudes $c_\mu(t)$ satisfy
\begin{eqnarray}
\dot{c}_\mu(t)=\sum_{\nu=1}^N M_{\mu\nu}c_\nu(t),
\end{eqnarray}
and  
\begin{eqnarray}
M_{\mu\nu} \equiv \frac{\Gamma}{2}(-F_{\mu\nu}+i2G_{\mu\nu}\delta_{\nu\neq\mu}).
\end{eqnarray} 
We further solve the above coupled equations by a similarity transformation, and diagonalize $M_{\mu\nu}$ with the eigenvalues $\lambda_l$ and eigenvectors $\hat{U}$, such that
\begin{eqnarray}
c_\mu(t)=\sum_{\nu,n} U_{\mu n}e^{\lambda_n t}U^{-1}_{n\nu}c_\nu(t=0),
\end{eqnarray}
where $c_\nu(t=0)$ denotes the initial condition of the system.\ 

We assume that the atoms are initially in one of the DM states $|\phi_m\rangle$, the state vector $|\Psi(t)\rangle$ is then time-evolved as $\sum_{m'=1}^N d_{m'}(t)|\phi_{m'}\rangle$ where the initial condition asserts that $d_{m'}(t=0)$ $=$ $\delta_{m'm}$.\ Using the relation of the coefficient transformation, \begin{eqnarray}
d_m = \frac{1}{\sqrt{N}}\sum_{\mu=1}^N c_\mu e^{-i\mathbf{k}\cdot\mathbf{r}_\mu}e^{-i\frac{2m\pi}{N}(\mu-1)},
\end{eqnarray} 
finally we derive the time evolution of the DM states \cite{Jen2016},
\begin{eqnarray}
d_m(t)=\sum_{n=1}^N v_n(m) e^{\lambda_n t}w_n(m),\label{evolve}
\end{eqnarray}
where 
\begin{eqnarray}
v_n(m)&\equiv&\sum_{\mu=1}^N \frac{e^{-i\mathbf{k}\cdot\mathbf{r}_\mu-i2m\pi(\mu-1)/N}}{\sqrt{N}}U_{\mu n},\label{weight1}\\
w_n(m)&\equiv&\sum_{\nu=1}^N U^{-1}_{n\nu}\frac{e^{i\mathbf{k}\cdot\mathbf{r}_\nu+i2m\pi(\nu-1)/N}}{\sqrt{N}}. \label{weight2}
\end{eqnarray}
The $v_n(m)$ is the inner product of $m$th DM state and $n$th eigenstate in $\hat{U}$, and $|v_n(m)|^2$ indicates how close DM states are to the eigen ones.\ We further use a normalized weighting of $|v_n(m)w_n(m)|^2$ as a measure of the portion for specific $\lambda_n$ that governs the time evolution of the DM states.\ Below we proceed to investigate the subradiance in optically thin (cubic) and thick (rectangular) atomic arrays, and locate the subradiant DM state that has longest lifetime for various lattice spacings and atom numbers.
\section{Time evolutions of the subradiant states}
Before we investigate the decay property of subradiant DM states, it is instructive to study the coupling strength of the induced dipole-dipole interaction in DM state bases.\ This will guide us to locate the lowest eigenvalues from the dependence of lattice spacings.\ We use a 3D atomic array of $N_x\times N_y\times N_z$ $=$ $2\times 2\times 4$ as an example, where $N$ $=$ $N_xN_yN_z$.\ We use a linear polarization of light ($\hat{d}$ $=$ $\hat{x}$) propagating along the long axis $\hat{z}$, and the coupling strength is defined as
\begin{eqnarray}
\Gamma_{m,m}=-2{\rm Re}\bigg[_{\rm 3D}\bigg\langle\phi_m\bigg|\sum_{\mu,\nu}M_{\mu\nu}\bigg|\phi_m\bigg\rangle_{\rm 3D}\bigg].
\end{eqnarray}

\begin{figure}[t]
\centering
\includegraphics[width=12cm,height=5.5cm]{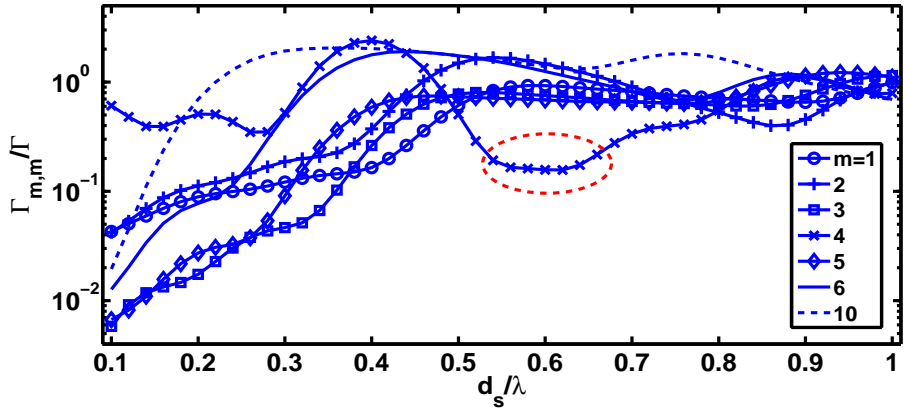}
\caption{(Color online) Coupling strengths $\Gamma_{m,m}$ of the DM states for various lattice spacings with $N_x\times N_y\times N_z$ $=$ $2\times 2\times 4$.\ Small coupling strengths appear at small lattice spacing $d_s$ $\approx$ $0.1-0.3\lambda$, and at plateau around $d_s$ $=$ $0.6\lambda$ (dash-circle) for the specific $m$ $=$ $4$ DM state.\ For larger lattice spacings, coupling strengths approach an order of $\Gamma$ indicating a regime of independent atoms. }\label{fig2}
\end{figure}

In figure \ref{fig2}, we show the coupling strengths for various DM states in a 3D atomic array of dimensions $N_x$ $\times$ $N_y$ $\times$ $N_z$ $=$ $2$ $\times$ $2$ $\times$ $4$, with $N$ $=$ $16$ atoms in total.\ It has no periodic dependence of lattice spacings as in one-dimensional (1D) finite \cite{Jen2016} or infinite atomic chain \cite{Nienhuisi1987} with a period of $\pi$ or $2\pi$ respectively.\ The reason is that the linearly increasing phases are imprinted on three perpendicular axes in our 3D case, such that equidistant shifted $m$th DM states do not follow the period of the sinusoidal functions in the dipole-dipole interactions.\ A new observation is relatively small coupling strength at the circled plateau where $d_s$ $\gtrsim$ $0.5\lambda$ for $m$ $=$ $4$ DM state.\ This puts the 3D DM states in advantage over 1D case to prepare a subradiant state since experimentally it is more feasible and less challenging to prepare and control the optical lattices at a spacing of $d_s$ $\gtrsim$ $0.5\lambda$.

\begin{figure}[t]
\centering
\includegraphics[width=12cm,height=6cm]{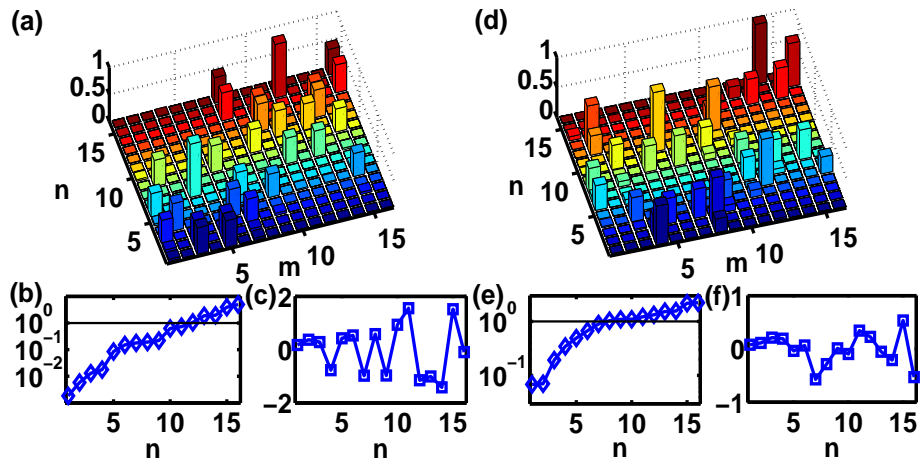}
\caption{(Color online) Normalized weightings of DM states $|\phi_m\rangle_{\rm 3D}$ on the eigenstates $|\phi_n'\rangle_{\rm 3D}$ for $N_x\times N_y\times N_z$ $=$ $2\times 2\times 4$.\ The weightings are observed in small groups of the eigenstates for both (a) $d_s$ $=$ $0.25$ and (d) $0.6\lambda$.\ Respectively the associated real ($-$Re$[2\lambda_n]/\Gamma$) and imaginary parts (Im$[2\lambda_n]/\Gamma$) of the eigenvalues are demonstrated in (b,e) and (c,f).\ The ascending order of (b) and (e) in logarithmic scale indicates the distribution of the superradiant (above $\Gamma$) and subradiant (below $\Gamma$) decay constants.\ A horizontal line is used to guide the eye for a natural decay constant (independent atoms regime).}\label{fig3}
\end{figure}

From figure \ref{fig2}, we later study specifically $d_s$ $=$ $0.25$ and $0.6\lambda$ which correspond to the lowest coupling strengths for the DM states
of $m$ $=$ $3(5)$ and $4$ respectively.\ These subradiant DM states will be demonstrated later to show the smallest decay rates (longest lifetimes) that could be potentially useful in quantum storage of light.\ Since in general the dipole-dipole interaction is long-ranged and has no exact analytical solutions of the eigenstates, the DM states we propose to prepare in the 3D atomic array would couple to several eigenstates that can be numerically derived.\ Therefore we plot the normalized weightings using Eqs. (\ref{weight1}) and (\ref{weight2}) in figure \ref{fig3}.\ 

The normalized weightings of $m$th DM states on the numerically derived eigenstates ($|\phi_n'\rangle_{\rm 3D}$) indicate how significant the eigenvalues $\lambda_n$'s dominate the decay process and also suggestively how close the DM states projected on the eigen ones.\ In our setting of rectangular array where $N_z$ $>$ $N_{x,y}$ in figure \ref{fig3}, we observe localized small groups of subspaces for the DM states.\ In figure \ref{fig3}(a), for example the pairs of $m$ $=$ $(3,5)$, $(2,6)$, and $(11,13)$ form almost closed subspaces of $n$ $=$ $(1,2)$, $(4,9)$, and $(8,11)$ respectively with the total subspace weightings of $95.8\%$, $99.9\%$, and $95.9\%$.\ Similar in figure \ref{fig3}(d) at a larger lattice spacing, a subspace of normalized weightings can be seen, which indicates an almost overlap with the eigenstates.\ The superradiant decay constants in (b) and (e) distribute at $n$ $>$ $10$ and $n$ $\gtrsim$ $8$ for small and large lattice spacings.\ Also more decay constants lie along $\lambda_n$ $=$ $\Gamma/2$ for a larger lattice spacing, which is reasonable since if $d_s$ keeps increasing, eventually the system approaches the regime of independent atoms.\ 

\begin{figure}[ht]
\centering
\includegraphics[width=12cm,height=5cm]{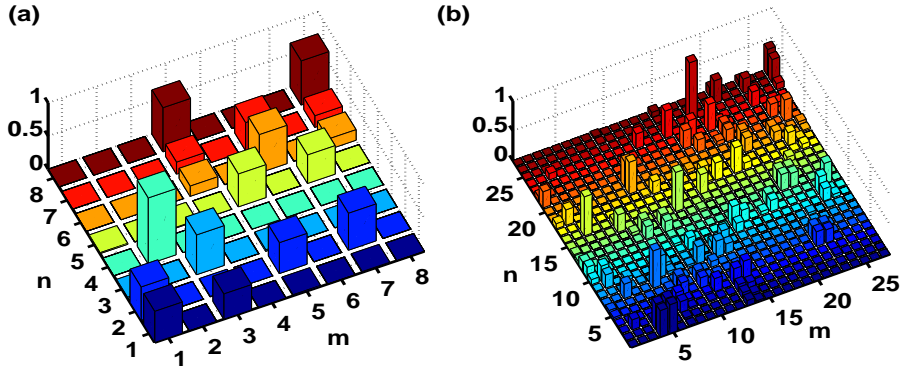}
\caption{(Color online) Normalized weightings of DM states $|\phi_m\rangle_{\rm 3D}$ on the eigenstates $|\phi_n'\rangle_{\rm 3D}$ for (a) $N_x\times N_y\times N_z$ $=$ $2\times 2\times 2$ and (a) $N_x\times N_y\times N_z$ $=$ $3\times 3\times 3$.\ In both cases, $d_s$ $=$ $0.25\lambda$.}\label{fig4}
\end{figure}

The feature of the localized subspaces in our rectangular setting disappears when we consider a cubic atomic array in figure \ref{fig4}.\ For demonstration we take $d_s$ $=$ $0.25\lambda$ as an example and consider $N$ $=$ $8$ and $27$ atoms respectively in (a) and (b).\ In (a) we clearly see the DM states of $m$ $=$ $1$, $3$, $5$, and $7$ are coupled with each other on the eigenstates of $n$ $=$ $1$, $2$, $3$, and $5$.\ No localized subspaces for even larger cubic atomic array are shown in (b).\ It suggests our proposed DM states are far from the eigenstates numerically derived in a cubic geometry.\ However, the DM states can still be prepared and assessed by phase imprinting such that the subradiant states with lowest decay constants in a cubic array can be utilized for quantum storage of single photons.\ Below we will investigate the time evolutions of subradiant states in rectangular and cubic atomic arrays, which can be observable in fluorescence measurements.

\begin{figure}[t]
\centering
\includegraphics[width=12cm,height=6cm]{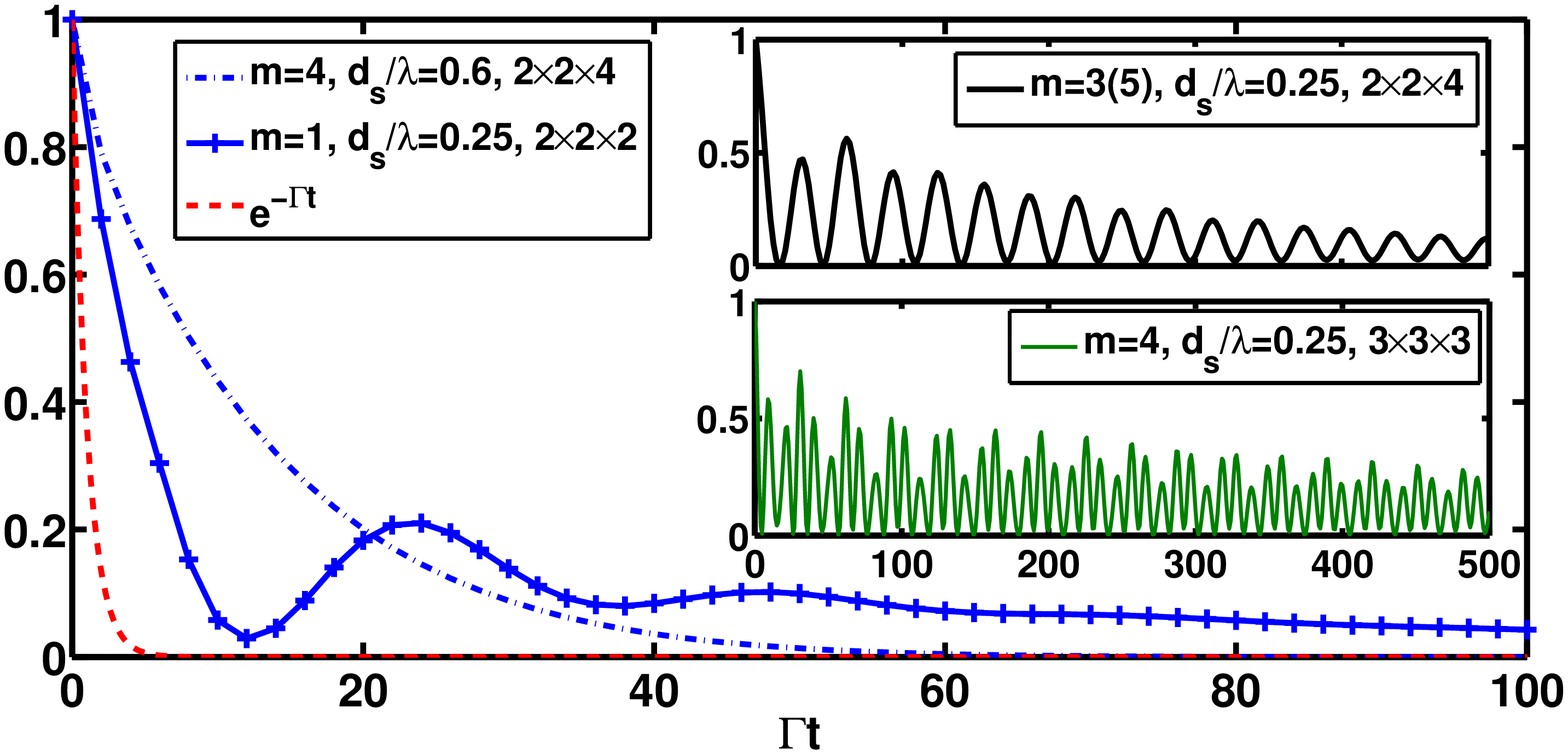}
\caption{(Color online) Time evolutions of cooperative single-photon subradiant states in 3D atomic arrays.\ The DM states with lowest decay constants are plotted for various lattice spacings $d_s$ $=$ $0.25$ or $0.6\lambda$ in a rectangular array of figure \ref{fig3} or a cubic one of figure \ref{fig4}.\ Atomic array with more atoms results in more reduction in the decay rate while the oscillatory structures result from the beating frequencies from two or more cooperative Lamb shifts of the eigenstates.\ The natural decay $e^{-\Gamma t}$ (dash) is shown as a reference for comparison and the insets share the same time dependence.}\label{fig5}
\end{figure}

The time-evolved DM states can be derived from Eq. (\ref{evolve}) where the probabilities $|d_m(t)|^2$ correspond to the measurements in fluorescence experiments.\ In figure \ref{fig5}, we plot the time evolutions of the DM states with the lowest decay constant from the settings in figures \ref{fig3} and \ref{fig4}.\ For the rectangular array we consider in figure \ref{fig3}, we find the lowest decay constant in the order of $5\times 10^{-3}\Gamma$ for $m$ $=$ $3(5)$ at $d_s$ $=$ $0.25\lambda$ (inset) and $7\times 10^{-2}\Gamma$ for $m$ $=$ $4$ at $d_s$ $=$ $0.6\lambda$.\ This suggests a lifetime of microseconds in the envelope of the decayed Rabi-like oscillation.\ For $N$ $=$ $27$ we consider in figure \ref{fig4}(b), the lifetime of $m$ $=$ $4$ DM state reaches an order of $26$ microseconds, a thousand-fold increase of the lifetime for free space decay of rubidium atoms ($26$ ns).\ With even more atoms, for example of $90$ atoms in an array of $N_x\times N_y\times N_z$ $=$ $3\times 3\times 10$ at a lattice spacing $d_s$ $=$ $0.25\lambda$, $m$ $=$ $22$ DM state reaches an even longer lifetime of $\sim 2$ milliseconds.

The beating frequency can also be observable in the fluorescence experiments as shown in figure \ref{fig5}.\ The major weightings of $m$ $=$ $3(5)$ on the pair of the eigenstates $n$ $=$ $1$ and $2$ for $N_x\times N_y\times N_z$ $=$ $2\times 2\times 4$ show one beating frequency of $0.2\Gamma$.\ Similarly for $m$ $=$ $1$ in an array of $2\times 2\times 2$, the major weightings of this DM state are on the eigenstates $n$ $=$ $1$ and $2$, giving an oscillatory but faster decay with a beating frequency of $0.26\Gamma$.\ This way we can determine the relative CLS from the fluorescence measurements of the DM states.\ More complicated beating structure is shown for a cubic array of $3\times 3\times 3$ for $m$ $=$ $4$ DM state which has up to three beating frequencies due to the spread out weightings on four major eigenmodes as can be seen in figure \ref{fig4}(b).

\section{Discussion and conclusion}
An issue of the DM states preparation protocol in figure \ref{fig1}(a) is the low efficiency of the initial superradiant state $|\phi_N\rangle_{\rm 3D}$.\ This superradiant state decays faster than single atomic decay ($\sim$ several nanoseconds for alkali metals), giving us an efficiency of $e^{-\Gamma_N\tau}$ where $\Gamma_N$ is the enhanced decay rate of the superradiant state we start with.\ To restore the efficiency we can use ultrashort light pulse to control and imprint the phases \cite{Scully2015} from linear Stark shift in a faster rate, thus with a higher efficiency.

Aside from the efficiency, the preparation of the DM states relies on the proper interaction time $\tau$ of the magnetic or electric field gradient which controls the imprinted phases.\ The inaccuracy of $\tau$ contributes to phase errors that influence the fidelity of the DM states we manage to prepare, which is twofold.\ One is the change of total weighting on the major subspace of the eigenstates and the other is the component of the weighting in this subspace.\ We note that the first change indicates a coupling outside the original subspace of the eigenstates, while both contribute to infidelity of the state preparation.\ Experimentally a few percent of phase error should be tolerable, and the fidelity of the proposed DM states do not deviate much from the perfect case \cite{Jen2016}.\ Therefore our preparation protocol of DM states should be resilient to the fluctuations of interaction time.

In conclusion, we propose a complete Hilbert space of cooperative single-photon states in a three-dimensional atomic array.\ These states can be realized in the two-level atoms confined in a three-dimensional optical lattice.\ Our proposal can provide a setting to investigate the subradiance by a proper phase imprint on the atoms.\ Cooperative Lamb shift can be also observable from fluorescence measurements in either rectangular or cubic arrays.\ For rectangular array, more localized and confined subspaces appear, indicating a close overlap of the DM states with the eigenstates.\ The lifetime of the manipulated DM states can be raised to an order of several milliseconds up to a hundred atoms, thus serving a robust quantum memory application.\ Our setting is also alternative to study the cooperative behavior in the square and kagome lattices \cite{Bettles2015} and the many-body long-range interactions in the alkaline-earth-metal atoms \cite{Olmos2013}.\

\section*{ACKNOWLEDGMENTS}
This work is supported by the Ministry of Science and Technology, Taiwan, with Grant No. MOST-101-2112-M-001-021-MY3.\ We also thank for the support of NCTS and the fruitful discussions with M.-S. Chang and Y.-C. Chen.

\end{document}